# The solar nitrogen abundance under the LTE assumption

Sameh I.S, M. M. Beheary, Abdelrazek M. K. Shaltout

Department of Astronomy and Meteorology, Faculty of Science, Al-Azhar University, 11884 Nasr City, Cairo, Egypt

## Abstract

The solar chemical abundances studied in considerable detail because of discrepant values of solar metallicity inferred from different indicators, where in the present work we have used 28 suitable atomic spectral solar nitrogen lines, relying on new equivalent widths measurements for determining the nitrogen abundance. Corresponding to 28 solar neutral nitrogen lines, we derived a solar abundance of nitrogen using the Holweger and Müller (1974) solar model and the measured equivalent widths from the literature. Our study reported the solar nitrogen abundance result to be of 8.05±0.09 using g$f$-values from Wiese and Fuhr (2007).

## 1. Introduction

Analysis of solar spectrum is the most basic ingredient for any solar abundance because of the large number of elements have been revised in several analyses (Asplund, Grevesse and Sauval 2005; Shaltout et al. 2013). Measurements of chemical abundances in the interstellar medium (ISM) are a powerful probe of galaxy evolution and star formation processes (Dave et al. 2012; Lilly et al. 2013; Lu et al. 2015; Belfiore et al. 2016).

In this research, we adopted solar spectrum of nitrogen which may be useful to determine the value of nitrogen abundance in the Sun. Based on the LTE assumption, the source function equal to the Planck function. Semi-empirical models were used in this case such as Gingerich (1969) and Mutschlecner and Keller (1972) solar models, and the most widely semi-empirical model used for solar abundances is the Holweger and Müller (1974) model. The nitrogen abundance can be determined from both atomic and molecular lines, although the forbidden N I lines are too weak in the solar spectrum to be measurable Grevesse et al. (1990). A preliminary nitrogen abundance was given in Asplund, Grevesse & Sauval (2005) using high excitation energy of N I lines and N H vibration rotation lines in the infrared while in the final analysis of Sauval et al. (2009) also N H pure rotation lines and a multitude of transitions from various bands of CN were considered in the framework of an improved 3D model. The non-LTE effects for N I are relatively small: −0.05 dex when ignoring H collisions as described in Caffau et al. (2009). In their own analysis of these very weak N I lines using their co5bold 3D solar model, Caffau et al. (2009) obtained a N I based abundance which is 0.05 dex higher than the corresponding value in Sauval et al. (2009), largely due to the selection of lines.

They included many lines that others considered it to be of dubious value due to it is suspected blends. This is also reflected in the radically different standard deviations for the N I lines in the two studies: $\sigma = 0.04$ and 0.12 dex, respectively. Sauval et al. (2009) found consistent abundances inferred from N I and NH solar lines, with very small dispersion in the NH results. The different CN bands also gave very similar results when adopting a self-consistently derived CN abundance using the 3D model. Unfortunately, Caffau et al. (2009) did not consider molecular lines and it is therefore not known how well their 3D model performs in achieving consistent results for atoms and molecules.

In this paper, we determined the photospheric nitrogen abundance using atomic data and Holweger-Müller solar model. The solar abundance of nitrogen among other elements was also determined by Asplund et al. (2005), using the 3D magneto hydrodynimical model. Both molecular and atomic lines were considered, this research found that the nitrogen abundance were found to be 7.73±0.05 and 7.85±0.08, for (N H &N I) respectively (including NLTE corrections). While Caffau et al. (2009) and Asplund et al. (2009) found the nitrogen abundance from NLTE with 3D models as 7.86 and 7.83, respectively. In the Sun, NLTE effects are generally found to be small, where the NLTE abundance correction is of the order of 0.05 dex (Rentzsch-Holm 1996).

## 2. Atomic Data

New equivalent widths for our selected 28 nitrogen solar lines are given in Table 1. The real difficulty for determine absolute equivalent width is to identify accurate flux at the beginning of departure, flux at the full width of half the maximum, flux at vertex and the height of absorption line, where these values have been estimated from Bass 2000 solar spectrum which is available at the web site (http://bass2000.obspm.fr/). Abundance calculations for any element will be correct due to the accurate selection of spectrum lines and atomic data parameters, we have chosen carefully the N I solar lines from NIST Atomic Spectra Database (Ralchenko 2005) in the range of 400−1200 nm. These lines were then checked using the solar spectrum identifications (Moore et al. 1966) for the 400−877 nm range, while the other lines are from Swensson et al. (1970) for the 900−1200 range. The equivalent width (W) of spectral line is defined as the total area of spectral line taking into account the continuum level of spectrum as the highest value. The method of MKK Book (1943) was used to calculate the equivalent widths of the important absorption lines.

**Table 1: Different equivalent widths (in mÅ) compared with those measured in this work.**

| Wavelength λ (Å) | Rentzsch-Holm (1996) | Grevesse et al. (1990) [total] | This work |
|---|---|---|---|
| 7442.3 | 2.6 | 3.3 | 3.16 |
| 7468.3 | 5.2 | 5.2 | 5.2 |
| 8184.78 | | | 4.1 |
| 8200 | 1.15 | 1.3 | 1.18 |
| 8216.3 | 8.6 | 8.6 | 8.6 |
| 8223.14 | 2.4 | 2.4 | 2.39 |
| 8242.39 | 3.9 | 4.2 | 4.13 |
| 8594 | | | 2 |
| 8629.24 | 4.5 | 6.35 | 4.74 |
| 8655.87 | 1.4 | 1.5 | 1.58 |
| 8683.4 | 7.8 | 8.7 | 8.2 |
| 8703.25 | | | 3.73 |
| 8711.7 | | | 4.54 |
| 8718.8 | 4.2 | 5.1 | 4.11 |
| 9045.878 | | 0.8 | 0.36 |
| 9392.79 | 9.7 | 11.6 | 10.5 |
| 9863.33 | | | 0.5 |
| 10105.13 | 1.8 | 1.8 | 1.4 |
| 10108.9 | 2.4 | 3.15 | 2.7 |
| 10112.5 | 3.5 | 3.5 | 3.3 |
| 10114.6 | 5.5 | 5.5 | 5.48 |
| 10507 | 1.4 | 1.4 | 0.61 |
| 10520.58 | 0.8 | 0.8 | 0.74 |
| 10539.57 | 3.2 | 3.2 | 3.11 |
| 10757.89 | 0.8 | 0.8 | 0.36 |
| 12381.65 | 1.35 | 2.25 | 1.14 |
| 12461.25 | 2.6 | 2.6 | 2.6 |
| 12469.62 | 5 | 5.5 | 4 |

**Note that:** Equivalent widths of solar N I lines of Rentzsch-Holm (1996) have typically the same values of Grevesse et al. (1990), but differ in the comparison with Grevesse et al. (1990) which have contributions of CN (taken as total values). This difference in two sets of equivalent width values may provide useful comparisons with regard to the values of equivalent widths. Equivalent widths of Grevesse et al. (1990), Rentzsch-Holm (1996), Asplund et al. (2009), Caffau et al. (2009) and this work have the same values of solar N I lines but differ at same lines, which have contributions of CN bands.

## 3. Analysis of solar nitrogen abundances

We used the WIDTH 9 FORTRAN code to determine the solar Nitrogen abundance depending on our selected lines, and the new determined equivalent widths (see, Table 1) at different sets of g$f$-values. We have updated the analysis which was made by Lambert (1978). All solar N I lines are presented in the spectrum of Delbouille et al. (1981) solar atlas. These lines are faint and many of them are blended with CN lines (Rentzsch-Holm 1996). We finally selected 28 N I lines belonging to the 3s-3p and 3p-3d transitions and adopted the Holweger & Müller (HM) solar model (see, Figure 2, for details). The accuracy in the equivalent widths of these N I lines is found to be +/-12%. Our equivalent widths are mostly consistent with Grevesse et al. (1990), Rentzsch-Holm (1996) and Caffau et al. (2009) for 3s-3p transition at the following lines (7468.30, 8216.30 and 8223.14 A). Also, our equivalent width values are typically in agreement with Grevesse et al. (1990) Rentzsch-Holm (1996) and Caffau et al. (2009) for the 3p3d transition for these lines (10112.50, 10114.60, 10520.58 and 12461.25 A). The uncertainty in gf-values is of +/-12%, when compared with the available published g$f$-values.

**Table 2: Solar nitrogen abundances with different equivalent width and g$f$-values.**

| Wavelength λ (Å) | $A_N$ 1 | $A_N$ 2 | $A_N$ 3 | $A_N$ 4 | g$f$ value by Rentzsch Holm | g$f$ values by Wise and Fuhr |
|---|---|---|---|---|---|---|
| 8683.4 | 8.183 | 8.026 | 8.026 | 8.053 | -0.051 | 0.105 |
| 8718.84 | 8.235 | 8.165 | 8.094 | 8.153 | -0.419 | -0.349 |
| 8216.32 | 8.198 | 8.072 | 8.035 | 8.072 | 0.012 | 0.138 |
| 8200.36 | 8.343 | 8.278 | 8.18 | 8.278 | -1.09 | -1.025 |
| 8242.39 | 8.159 | 8.071 | 8.013 | 8.096 | -0.36 | -0.272 |
| 8223.14 | 7.951 | 7.836 | 7.814 | 7.834 | -0.39 | -0.275 |
| 7468.31 | 8.381 | 8.167 | 8.238 | 8.167 | -0.397 | -0.183 |
| 7442.3 | 8.212 | 8.04 | 8.075 | 8.134 | -0.573 | -0.401 |
| 9392.79 | 8.186 | 8.202 | 8.019 | 8.247 | 0.316 | 0.3 |
| 8629.24 | 8.074 | 8.066 | 7.915 | 8.093 | 0.069 | 0.077 |
| 8655.87 | 8.215 | 8.204 | 8.066 | 8.259 | -0.63 | -0.619 |
| 10114.6 | 8.254 | 8.228 | 8.114 | 8.226 | 0.751 | 0.777 |
| 10112.5 | 8.183 | 8.149 | 8.054 | 8.12 | 0.588 | 0.622 |
| 10108.9 | 8.201 | 8.161 | 8.071 | 8.198 | 0.403 | 0.443 |
| 10105.13 | 8.247 | 8.212 | 8.122 | 8.095 | 0.2 | 0.235 |
| 10539.57 | 8.221 | 8.221 | 8.099 | 8.207 | 0.53 | 0.53 |
| 10507 | 8.111 | 8.242 | 8 | 7.86 | 0.25 | 0.118 |
| 10520.58 | 8.145 | 8.081 | 8.04 | 8.159 | -0.04 | 0.024 |
| 10757.89 | 8.166 | 8.475 | 8.058 | 8.122 | -0.08 | -0.389 |
| 12381.65 | 7.829 | 7.971 | 7.859 | 8.099 | 0.32 | 0.178 |
| 12469.62 | 8.233 | 8.233 | 8.163 | 8.124 | 0.61 | 0.61 |
| 12461.25 | 8.072 | 8.118 | 8.018 | 8.118 | 0.451 | 0.405 |
| 8184.87 | | | | 8.158 | | -0.305 |
| 8594 | | | | 8.082 | | -0.335 |
| 8703.25 | | | | 8.06 | | -0.31 |
| 8711.703 | | | | 8.083 | | -0.233 |
| 9045.878 | | | | 7.854 | | 0.439 |
| 9863.33 | | | | 7.818 | | 0.08 |
| | 8.18+/-0.105 | 8.12+/-0.106 | 8.05±0.09 | 8.10+/-0.120 | | |

Note: Column 2: refers to g$f$-values and equivalent width from Rentzsch-Holm (1996). Column 3: refers to g$f$-values and equivalent width from Grevesse et al. (1990). Column 4: refers to g$f$ values from Rentzsch-Holm (1996) and our equivalent widths. Column 5: refers to g$f$-values from Grevesse et al. (1990) and our equivalent widths. The g$f$-values from Wise and Fuhr (2007) are the same values of Grevesse et al. (1990).

**Table 3: Some contributions for the nitrogen abundance in the Sun as compared with the present work.**

| Source | $A_N$ |
|---|---|
| Kurucz et al. (1966) | 8.05±0.09 |
| Peytremann (1975) | 8.00±0.14 |
| Grevesse *et al.* (1990) | 8.01±0.10 |
| Hibbert *et al.* (1991) | 8.02±0.15 |
| Opacity project (1992) | 8.05±0.09 |
| Rentzsch-Holm (1996) | 8.05±0.09 |
| This work | 8.05±0.09 |

In this table, we have compared our abundance value and previous contributions results. This analysis was calculated under the assumption of LTE while previous contributions had been calculated under the NLTE assumption.

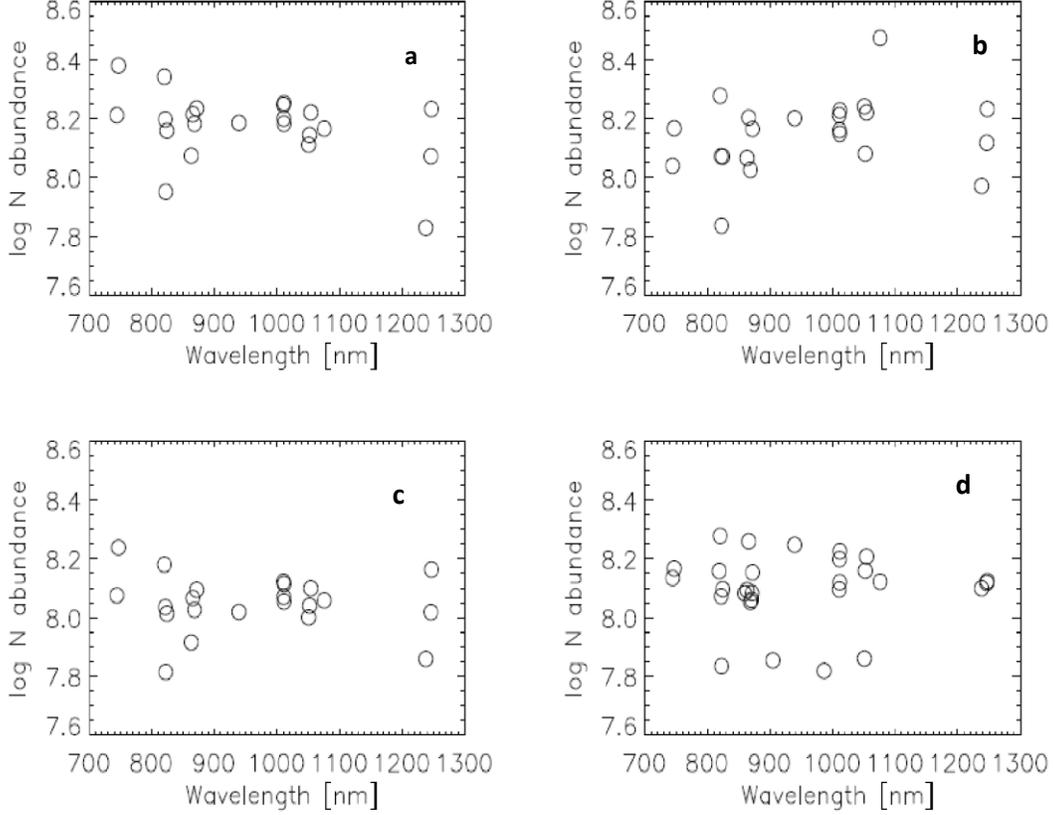

Fig. 1: The nitrogen abundance in the Sun calculated with different sets of *gf*-values and different values of equivalent widths obtained with the Fortran Width9 code under the LTE assumption: **a** g*f*-values and equivalent widths from Rentzsch-Holm (1996), **b** g*f* values and equivalent width from Grevesse et al. (1990), **c** g*f*-values from RentzschHolm (1996) and our equivalent widths, **d** g*f*-values from Grevesse et al. (1990) and our equivalent widths.

## 4. LTE theoretical Temperature model

In the present work, the most widely semi-empirical solar model used for deriving abundance of nitrogen is the Holweger and Müller (1974, hereafter, HM) model, while our theoretical model is obtained under the assumption of LTE in radiative-convective equilibrium, where convection is treated in a mixing-length approach. For more details, our LTE theoretical temperature model is derived with effective temperature ($T_{eff}$ = 5777 Kelvin), the surface gravity (cm/sec2) (log g = 4.4377) and the metallicity (log [M/H] = 0.0). The solar photospheric abundances of all elements are adopted from Grevesse and Sauval (1998). The microturbulent velocity in the line opacity with 1.0 km s$^{-1}$ is taken into account. We use Atlas9 Fortran code written by kurucz (1993) which is widely used to produce LTE one-dimensional temperature models.

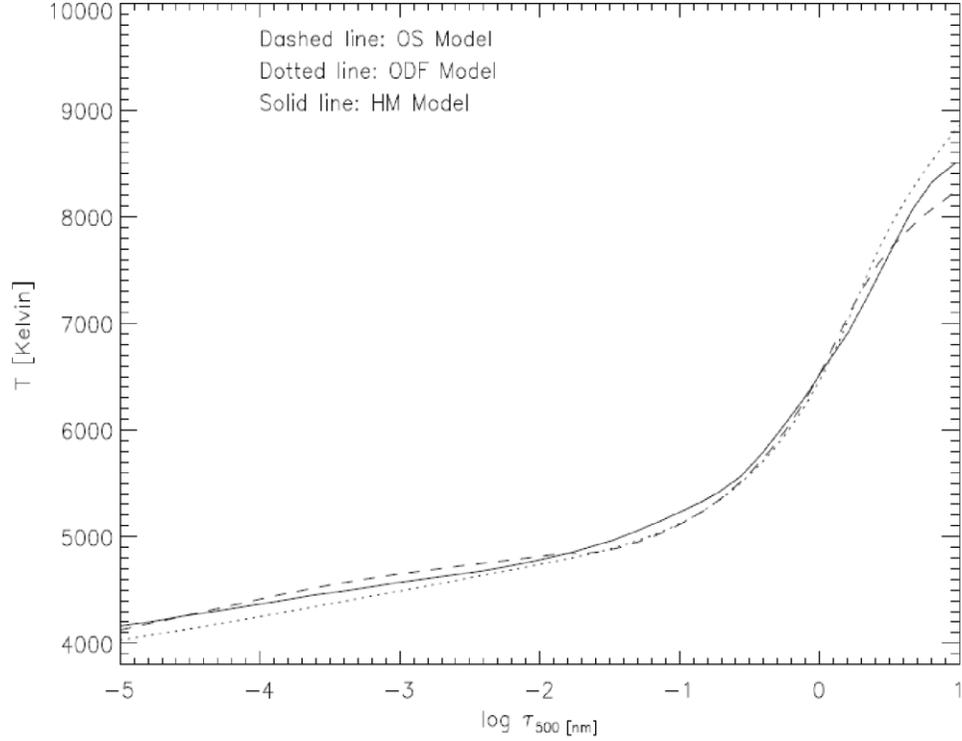

Fig. 2: Temperature stratifications of Opacity Distribution Function (ODF) (dotted line) Opacity sampling (dashed line) and HM (solid line) solar models.

## 5- Conclusion

Our study was done by selecting 28 solar nitrogen lines, then we have measured new equivalent widths for these lines using Bass 2000 Atlas of solar spectrum. In this study, we have used the WIDTH 9 FORTRAN code to calculate the solar nitrogen abundance based on our selected lines with new equivalent widths and g$f$-values from Rentzsch-Holm (1996), and we found that the abundance of nitrogen equals to 8.05±0.09 and while for the Grevesse et al. (1990) g$f$-values the abundance is 8.10±0.120. Also, we derive the solar nitrogen abundance using g$f$ values and equivalent widths from Rentzsch-Holm (1996), and it is reported as 8.18 ±0.105, while we used the g$f$-values and equivalent widths from Grevesse et al. (1990). Hence our calculations give 8.12 ±0.106. Our study achieved that, as a conclusion the solar nitrogen abundance is 8.05±0.09, which the same result obtained by Rentzsch-Holm (1996), Kurucz *et al.* (1966) and Opacity project (1992), while Peytremann (1975) gave the nitrogen abundance as 8.00±0.14. On the other hand, Grevesse *et al.* (1990) found the nitrogen abundance to be 8.01±0.10 while Hibbert et al. (1991) derived the nitrogen abundance by the value 8.02±0.15.